# Model based IVHM system for the solid rocket booster

Dmitry G Luchinsky and Vyatcheslav V. Osipov
NASA Ames Research Center, MS 269-3, Moffett Field, CA, 94035,USA and
Mission Critical Technologies Inc., 2041 Rosecrans Ave., Suite 225 El Segundo, CA 90245
Vadim N. Smelyanskiy, Dogan A. Timucin, and Serdar Uckun
NASA Ames Research Center, MS 269-2, Moffett Field, CA, 94035,USA
650-604-2261
Vadim.N.Smelyanskiy@nasa.gov

*Abstract*— We report progress in the development of a model-based hybrid probabilistic approach to an on-board IVHM for solid rocket boosters (SRBs) that can accommodate the abrupt changes of the model parameters in various nonlinear dynamical off-nominal regimes. The work is related to the ORION mission program. Specifically, a case breach fault for SRBs is considered that takes into account burning a hole through the rocket case, as well as ablation of the nozzle throat under the action of hot gas flow. A high-fidelity model (HFM) of the fault is developed in FLUENT in cylindrical symmetry. The results of the FLUENT simulations are shown to be in good agreement with quasi-stationary approximation and analytical solution of a system of one-dimensional partial differential equations (PDEs) for the gas flow in the combustion chamber and in the hole through the rocket case.

The low-dimensional performance model (LDPM) of the fault is derived by integrating a set of one-dimensional PDEs along the axis of the rocket. The LDPM is used to build a model-based fault diagnostic and prognostic (FD&P) algorithm for the case breach fault. In particular, two algorithms are introduced. The first algorithm is based on the self-consistent algorithm that solves the LDPM in a quasi-adiabatic approximation, when the pressure and density follow adiabatically dynamics of the propellant burning, melting and burning of the metal case, ablation and erosion of nozzle and insulator. The second algorithm is based on the dynamical inference method [1]-[3] of the system of stochastic differential equations of the LDPM.

The parameters of the HFM model and of the LDPM are tuned to reproduce the results of recent experiments of the rocket firing with the case breach fault in the forward closure. The FD&P is then applied to illustrate real-time diagnostics of the model parameters and prognostics of the SRB internal ballistics. All the algorithms discussed in this paper were verified using experimental data as will be discussed elsewhere.

The accuracy of the algorithm and the possibility of its application to FD&P for other SRB fault modes are discussed.

### TABLE OF CONTENTS

### NOMENCLATURE

$\rho$ = gas density
$p$ = gas pressure
$T$ = gas temperature
$u$ = gas velocity
$v_m$ = velocity of metal melting front
$v_n$ = velocity of nozzle ablation front
$c$ = sound velocity
$M$ = Mach number, $M = u/c$
$c_V$ = specific heat for constant volume
$c_p$ = specific heat for the constant pressure
$\gamma$ = ration of specific heats $\gamma = c_P/c_V$
$l$ = perimeter of propellant cross-section
$e_t$ = total energy of the gas
$h_t$ = total enthalpy of the gas
$l_h$ = perimeter of hole cross-section
$r_n$ = radius of nozzle throat
$s_t$ = cross section of the nozzle throat
$r_h$ = minimum radius of leak hole
$s_h$ = cross-section of hole throat
$r_{met}$ = radius of hole in metal case
$r_i$ = radius of hole in insulator layer
$L$ = length of the propellant grain
$L_0$ = typical length equal to 1m
$S_b$ = total area of the burning surface
$F_N$ = normal thrust
$F_h$ = additional thrust produced by hole gas flow
$r_b$ = burning rate of solid propellant
$n$ = exponent for burning rate of the propellant
$\rho_p$ = density of the solid propellant
$H_p$ = combustion heat of the solid propellant
$Q$ = heat flow from the gas to the walls of the hole
$S$ = cross-section of the combustion chamber
$f_{tr}$ = surface friction force

$T_m$ = melting temperature point
$T_a$ = critical temperature of the nozzle ablation
$T_c$ = temperature of metal case far from hole
$c_m$ = specific heat of case metal
$q_{ins}$ = latent heat of insulator ablation
$q_m$ = latent heat of metal melting
$c_n$ = specific heat of nozzle material
$q_n$ = specific heat of ablation of insulator layer
$\rho_m$ = density of case metal
$\rho_n$ = density of nozzle material
$k$ = the thermal conductivity
$\mu$ = dynamical viscosity of hot gas
Pr = the Prandtl number, Pr=$\mu C_p/k$
$h$ = subscript for gas parameters in the hole
$N$ = subscript for parameters in normal regime
$0$ = subscript for stagnation values of gas parameters

## 1. INTRODUCTION

New generation of the heavy lift vehicles for space exploration, including ORION mission, requires development of the new IVHM systems with the overarching goal of safer and more reliable flights. The difficulties in the development of an on-board IVHM for SRBs are the facts that: (i) internal hydrodynamics of SRBs is highly nonlinear, (ii) there is a number of failure modes that may lead to abrupt changes of SRBs parameters, (iii) the number and types of available sensors are severely limited, and (iv) the recovery time is typically a few seconds. These difficulties dictate the model based approach to the IVHM of SRBs that minimizes the number of "misses" and "false alarms" [4] by utilizing deep understanding of the physical processes underlying the nominal and off-nominal regimes of SRBs.

Indeed, theoretical analysis shows that many fault modes leading to the SRBs failures [5]-[7], including combustion instabilities [8]-[10], bore choking [11]-[13], and case breach [5], [14], have unique dynamical features in the time-traces of pressure and thrust. The corresponding expert knowledge can be incorporated into on-board FD&P within the novel Bayesian inferential framework [1]-[4] allowing for faster and more reliable identification of the nominal and off-nominal regimes of SRB operation in real time.

However, the development of such an inferential framework in practice is a highly nontrivial task since the internal dynamics of gas flow in the SRBs is essentially nonlinear and results from the interplay of a large number of complex nonlinear dynamical phenomena on the propellant and insulator and metal surfaces, in the gas flow, and in the nozzle. On-board FD&P, on the other hand, can only incorporate low-dimensional performance models (LDPM) of the pressure and thrust dynamics of SRBs. The derivation of the corresponding LDPMs and its validation in high-fidelity simulations and experimental rocket firing tests are important ingredients of the development of the FD&P.

A very important issue of the development of on-board IVHM system for SRBs is its ability to accommodate the abrupt changes of the model parameters in various nonlinear dynamical off-nominal regimes. The difficulties in building model-based hybrid probabilistic algorithms for on-board IVHM stem from the fact that at present there are no general inferential frameworks for nonlinear dynamical systems with mixed binary and continuous parameters.

In the present paper we address both problems in the context of derivation and testing of a LDPM for on-board FD&P. Specifically, we introduce a high-fidelity model of the case breach fault, derive a corresponding LDPM, and compare the results of the simulations of both models. The LDPM is then applied to analyze two situations motivated by recent ground firing tests. In the first example, we assume that the hole in the metal case appears suddenly (perhaps due to an interaction with an external object), the hole in the metal is larger than the hole in the insulator and the dynamics of the fault is determined by the dynamics of the ablation of the insulator. We show that, in this case, FD&P can be

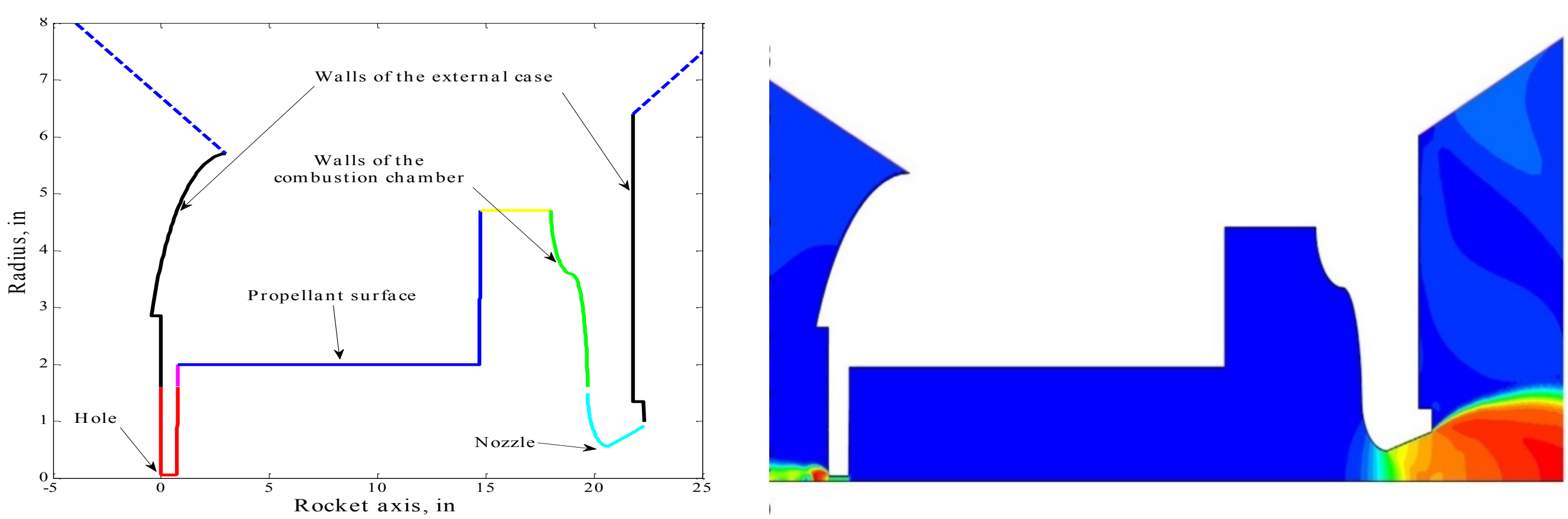

**Figure 1 (left) Geometry of the model surfaces in FLUENT. Note that the hole wall, propellant surface wall, and the nozzle wall are deforming. The dashed blue lines on the left and right hand side show boundaries of the ambient regions. (right) Velocity distribution generated by the FLUENT model for t = 0.34 sec.**

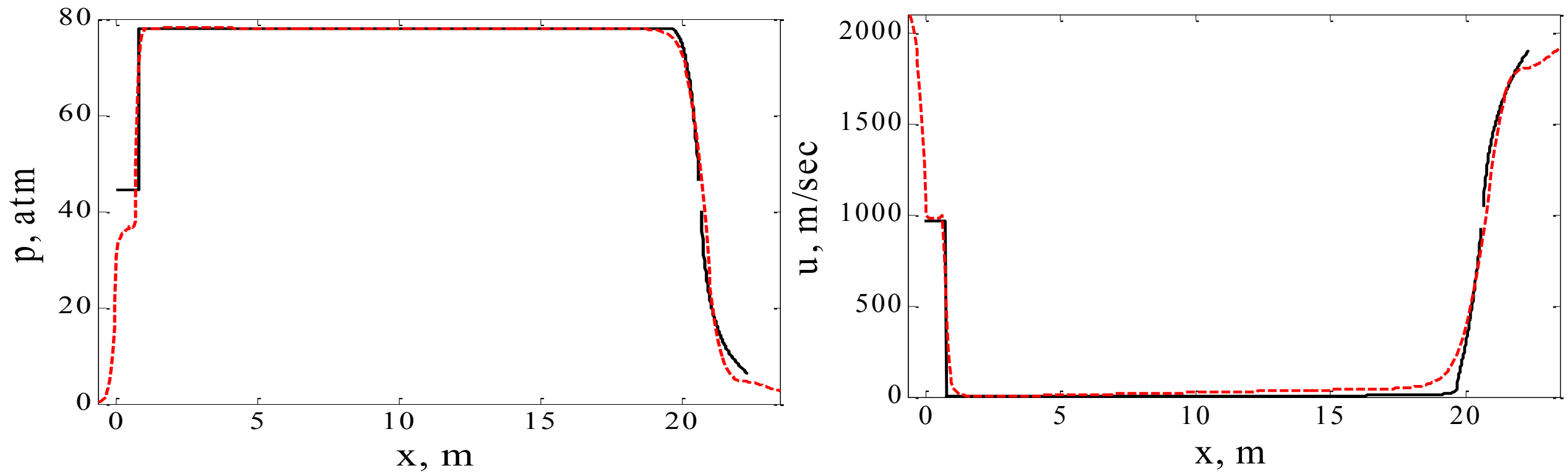

**Figure 2 Axial velocity (left) and pressure (right) profiles generated by the FLUENT model for t = 0.05 sec (black solid line) as compared to the analytical solutions (red dashed lines) given by the Eqs (7)-(8).**

performed using a self-consistent iterative algorithm that avoids numerical integration of the LDPM. It is assumed in this case that the geometry of the fault is simple, so that the dynamics of the fault is governed by equations for the ablation of the nozzle and insulator, and melting and burning of the hole in the metal case based on Bartz' approximations [20]-[23] .

In the second example we consider the situation when the hole in the metal case develops more slowly perhaps due to the crack in the insulator. In this case the dynamics is governed by the melting of the metal. The geometry of the fault in this example is more complicated and is motivated by the recent ground tests. In this example the fault dynamics deviates substantially from the one given by the Bartz' approximation. Accordingly a more general algorithm for FD&P is introduced.

We show that the introduced algorithms can accommodate abrupt changes in the model parameters of essentially nonlinear dynamics of the gas flow in SRBs. The applicability of the results to different SRB fault modes is discussed.

## 2. CASE BREACH MODEL

In this section we develop both axisymmetric FLUENT [25] HMF of the case breach fault and a set of PDEs that describe this fault in 1D approximation.

To model a case breach fault (see the sketch in Figure 1) we assume that the gas leak through the hole in the case is small compared to the gas flow through the rocket nozzle. An analysis of the fault caused by a small gas leak is the most important one from the point of view of the development of prognostics and diagnostic for a SRB. An axisymmetric HFM of this fault is built in FLUENT. The geometry of the model surfaces is shown in Figure 1. This model takes into account the nozzle ablation and the melting of the metal in the hole through the rocket case. Even more importantly, it models the dynamics of the burning area as a given function of the burn web distance. These features allow us to reproduce accurately experimentally observed time-traces of the rocket and gas flow parameters. An example of the velocity distribution generated by this model is shown in Figure 1.

The dynamics of the distributions of the gas flow parameters obtained with the FLUENT model can now be applied to validate low-dimensional performance model (LDPM) of SRBs. To do so we notice that in axisymmetric case one can use one-dimensional approximation for both gas flow in the combustion chamber and gas flow in the hole[1]. To describe gas flow in the combustion chamber in 1D approximation we can use a set of equations for the dynamics of mass, momentum, and the energy of the gas, introduced in our earlier work [3],[4].

$$\begin{cases} \partial_t \left(S\rho\right) = -\partial_x \left(S\rho u\right) + \rho_p r_b l + A_1\xi_1(t), \\ \partial_t \left(S\rho u\right) = -\partial_x \left(S\rho u^2\right) - S\partial_x p + A_2\xi_2(t), \\ \partial_t \left(S\rho e_t\right) = -\partial_x \left(S\rho h_t u\right) + H\rho_p r_b l + A_3\xi_3(t). \end{cases} \qquad (1)$$

Here $\partial_t \equiv \partial/\partial t$, $\partial_x \equiv \partial/\partial x$, $e_t = c_V T + u^2/2$, $h_t = c_P T + u^2/2$ are the total energy and total enthalpy of the gas flow in the combustion chamber, $\xi_i(t)$ are white Gaussian noises with intensities $A_i$. The boundary conditions for Eqs (1) assume the continuity conditions at the nozzle inlet and the stagnation values for the gas parameters at the point where the Mach number is zero in the combustion chamber. Eqs. (1) represent a modification of the well-known model of Salita [15]. Note also that in general case the gas dynamics is stochastic and the noise terms $A_i\xi_i(t)$ should be included into the equations.

We now extend this mode by coupling the gas dynamics in the combustion chamber to the gas flow in the hole. The corresponding set of PDEs

3

[1] Note that the same approximation is valid for arbitrary geometry of the fault in the limit small gas leak when the perturbation of the gas flow in the combustion chamber is small.

$$
\begin{cases}
\partial_t\left(s_h\rho_h\right) = -\partial_x\left(s_h\rho_h u_h\right), \\
\partial_t\left(s_h\rho_h u_h\right) = -\partial_x\left(s_h\rho_h u_h^{2}\right) - s_h\partial_x p_h - f_{fr} l_h, \\
\partial_t\left(s_h\rho_h e_t^{h}\right) = -\partial_x\left[s_h\rho_h u_h h_t^{h}\right] - Q_h l_h,
\end{cases} \quad (2)
$$

resembles Eqs. (1). The important difference, however, is that we neglect mass flow from the walls of the hole. Instead Eqs. (2) include the term  that describes the heat flow from the gas to the hole walls. The boundary conditions for this set of equations assume ambient conditions at the hole outlet and the continuity equation for the gas flow in the hole coupled to the sonic condition at the hole throat. The value of $Q_h$ is presented in the Appendix. The dynamics of the gas flow in the nozzle is described by a set of equations similar to (2) and can be obtained from this set by substituting subscript "n" for subscript "*h*".

The sets of PDEs for the gas dynamics in the combustion chamber (1), hole (2), and nozzle area are coupled with the equation for the steady burning of the propellant

$$
\frac{dR}{dt} = r_b\left(p\right) = r_c\left(\frac{p}{p_c}\right)^{n} = ap^{n}, \quad (3)
$$

and equations for the nozzle ablation

$$
\frac{dr_n}{dt} = v_n\left(\frac{\rho v}{\left(\rho v\right)_{nt}}\right)^{1-\beta}\left(\frac{r_n(t,x)}{r_{n0}}\right)^{-\beta}\left(\frac{T-T_a}{T_{*}-T_a}\right) \quad (4)
$$

and melting of the metal walls of the hole in the rocket case

$$
\frac{dr_h}{dt} = v_m\left(\frac{\rho v}{\left(\rho v\right)_{ht}}\right)^{1-\beta}\left(\frac{r_h(t,x)}{r_{h0}}\right)^{-\beta}\left(\frac{T-T_m}{T_{*}-T_m}\right). \quad (5)
$$

Velocities of the melting $v_m$ and ablation $v_n$ fronts are given in the Appendix.

It is further assumed that throughout the combustion chamber and in the hole the following equation of a perfect gas holds

$$
\frac{p}{\rho} = (C_p - C_V)T = \frac{p_0}{\rho_0}\left(\frac{T}{T_0}\right) = \frac{c_0^2}{\gamma}\left(\frac{T}{T_0}\right). \quad (6)
$$

In a quasi-steady regime Eqs (1)-(6) can be integrated analytically [4], [14] (cf. with numerical solution in [16]) leading to the following set of equations in the combustion chamber

$$
\begin{aligned}
\rho &= \rho_0\left(1+\frac{\gamma+1}{2}\frac{u^2}{c_0^2}\right)^{-1}, \\
p &= p_0\left(1+\frac{\gamma+1}{2}\frac{u^2}{c_0^2}\right)^{-1}\left(1-\frac{\gamma-1}{2}\frac{u^2}{c_0^2}\right)^{-1}, \\
v &= u_L\frac{x}{L}\left(1+\frac{3(\gamma+1)-2n\gamma)}{6c_0^2}u_L^2\frac{x^2}{L^2}\right).
\end{aligned} \quad (7)
$$

In the nozzle and the hole areas the analytical solution can be written in a well-known (see e.g. [17]) form

$$
\begin{aligned}
&p = p_0 M_{fact}^{\frac{\gamma}{\gamma-1}}, \quad \rho = \rho_0 M_{fact}^{\frac{1}{\gamma-1}}, \\
&T = T_0 M_{fact}, \quad \mathrm{M}_0\left(1-\frac{\left(\gamma-1\right)}{2}M_0^2\right)^{\frac{1}{\gamma-1}} = \frac{s_t}{\Gamma S},
\end{aligned} \quad (8)
$$

where $M_{fact} = \left(1-\left(\gamma-1\right)M_0^2/2\right)$. The results of the analytical solution are in good agreement with the results of the CFD simulations in FLUENT as shown in Figure 2 for time $t$ = 0.05 sec. Similar results are obtained for later times, demonstrating a good agreement between theory and simulations of time variation of quasi-steady axial distributions of the flow parameters. The good agreement between the analytical results and the time variation of the axial distributions of the gas flow parameters in FLUENT is the first step in the validation of the low-dimensional performance model of the fault derived in the next Section.

## 3. LOW-DIMENSIONAL PERFORMANCE MODEL

It follows from the results of analytical calculations and simulations of the FLUENT HFM that $M^2 = v^2/c_0^2 = 1$ is small everywhere in the combustion chamber. Furthermore, the equilibration of the gas flow parameters in the combustion chamber occurs on the time scale of the order of $t = L/c \approx 6$ msec. As a result, the distribution of the flow parameters follows adiabatically the changes in the rocket geometry induced by the burning of the propellant surface, nozzle ablation and metal melting in the hole through the case. It is, therefore, possible to derive the LDPM of the case breach fault by integrating Eqs (1), (2) (see [3], [4], [14] for further details) and combining the result of integration with Eqs (3)-(6) to obtain the following set of six ODEs

$$
\begin{cases}
\dot{\rho} = -\dfrac{c_0 \Gamma s_{et}}{V r_b} \rho \sqrt{\dfrac{p}{\rho}} + \dfrac{S_b}{V}\left(\rho_p - \rho\right) p^n + a_1 \xi_1(t), \\
\dot{p} = -\dfrac{c_0 \gamma \Gamma s_{et}}{V r_b} p \sqrt{\dfrac{p}{\rho}} + \dfrac{S_b}{V}\left(\gamma \rho_p - p\right) p^n + a_2 \xi_2(t), \\
\dot{R} = a p^n, \quad \dot{V} = S_b \dot{R} = S_b a p^n, \\
\dot{r}_t = v_{n0} p^{1-\beta_t} \left(\dfrac{r_t}{r_{t0}}\right)^{-\beta_t}, \\
\dot{r}_h = v_{h0} p^{1-\beta} \left(\dfrac{r_h}{r_{h0}}\right)^{-\beta} \left(\dfrac{T_h - T_m}{T_* - T_m}\right).
\end{cases}
\tag{9}
$$

Here the following dimensionless variables are used

$$
\begin{aligned}
& p \to \frac{p_0}{p_m}, \rho \to \frac{\rho_0}{\rho_m}; t \to \frac{t r_b\left(p_m\right)}{L_0}, r_t \to \frac{r_t}{L_0} \\
& F \to \frac{F}{L_0^2}, V \to \frac{V}{L_0^3}, r_{ht} \to \frac{r_{ht}}{L_0}, R \to \frac{R}{L_0}, s_{et} \to \frac{s_{et}}{L_0^2}
\end{aligned}
, \tag{10}
$$

where subscript $m$ refers to maximum reference values of the pressure and density. Important novel features of the case breach LDPM derived above are the following:

(1) The burning area of the propellant $S_b$ is calculated using a given design curve $S_b = f(R)$ that relates $S_b$ to the burn web distance $R$.
(2) The dynamics of the case breach fault is taken into account by including the last equation for melting the metal walls of the hole through the rocket metal case.
(3) The effective nozzle area $s_{et} = \pi r^2_{ht} + \pi r^2_t$ is introduced in the first two equations that take into account the effect of the fault dynamics on dynamics of the flow parameters at the stagnation point.
(4) The dynamics of the volume of the combustion chamber is taken into account in the equation 4 of the set of Eqs. (9).
(5) The dynamics of the nozzle ablation is taken into account in the last equation of the set.

In the general case the set of Eqs (9) may also include an equation for the dynamics of the hole in the insulator layer

$$
\dot{r}_t = v_{i0} p^{1-\beta_i} \left(\frac{r_i}{r_{i0}}\right)^{-\beta_i} \tag{11}
$$

These features allow us to model realistic changes in the rocket geometry with arbitrary thrust curves.

In the following sections we discuss probabilistic algorithms for an on-board SRB FD&P based on the model introduced above. These algorithms should satisfy a few very strict requirements mentioned in the introduction. In particular, these algorithms should accommodate very short time-windows for the diagnostics and prognostics (of the order of a few seconds) and abrupt changes of the model parameters reflecting sudden transitions from nominal to off-nominal regimes. As the main input, they use time-traces of stagnation pressure and thrust. These algorithms should allow for discrimination between multiple fault modes of SRBs.

In particular, we consider two algorithms. The first algorithm is based on the self-consistent algorithm that solves the LDPM in a quasi-adiabatic approximation, when the pressure and density follow adiabatically dynamics of the propellant burning, melting of the metal case, and nozzle and insulator ablation. The second algorithm is based on the integration of the system of stochastic differential equations of the LDPM.

## 4. INFERENCE OF THE MODEL PARAMETERS

Note that effect of the case breach fault on the dynamics of the internal gas flow in SRBs is reduced to the effective modification of the nozzle throat area as explained above. It is, therefore, possible to infer SRB parameters using a Bayesian framework introduced in our earlier work [1]-[4], [14], [30] for an analysis of the overpressure faults due to the changes of the nozzle throat area. In particular, it was shown [4], [14], [30] that this algorithm can accommodate sudden changes of the model parameters and, therefore, is suitable for developing of the hybrid probabilistic IVHM of SRBs.

Here we briefly reproduce earlier the results related to the analysis of the abrupt changes of the model parameters. The dynamics of the LDPM (9) can be in general presented in the form of Euler approximation of the set of ODEs on a discrete time lattice $\{t_k = hk; k = 0,1,...,K\}$ with time constant $h$

$$x_{k+1} = x_k + hf(x_k^* \mid c) + \hat{\sigma}\sqrt{h}z_k, \qquad (12)$$

where $z_k = \frac{1}{\sqrt{h}}\int_{t_k}^{t_k+h} \xi(t)dt$, $x_k^* = \frac{x_k + x_{k+1}}{2}$, $\boldsymbol{x_k}$ = $\{p,\rho,R,V,r_h,r_t,r_i\}$ is $L$-dimensional state of the system (9), (11), $\sigma$ is in our case a diagonal noise matrix with two first non-zero elements $a_1$ and $a_2$, $\boldsymbol{f}$ is a vector field representing the rhs of this system, and $c$ are parameters of the model. Given prior distribution for the unknown model parameters in a form of Gaussian distribution we can apply our theory of Bayesian inference of dynamical systems [1]-[4] to

$$D_{ij} = \frac{h}{K}\sum_{k=0}^{K-1}(\dot{x}_k - f(x_k;c))_i(\dot{x}_k - f(x_k;c))_j \qquad (13)$$

$$c'_l = (A)_{ml}^{-1} w_m, \qquad (14)$$

$$w_m = h\sum_{k=0}^{K-1}\left[\sum_{n,n'=1}^{L} U'_{mn}(t_k)\, D_{nn'}^{-1}\dot{x}_{n'}(t_k) - \frac{\mathrm{v}_m}{2}\right] \qquad (15)$$

$$A_{ml} = h\sum_{k=0}^{K-1}\left[\sum_{n,n'=1}^{L} U'_{mn}(t_k) D_{nn'}^{-1} U_{n'l}(t_k)\right]. \qquad (16)$$

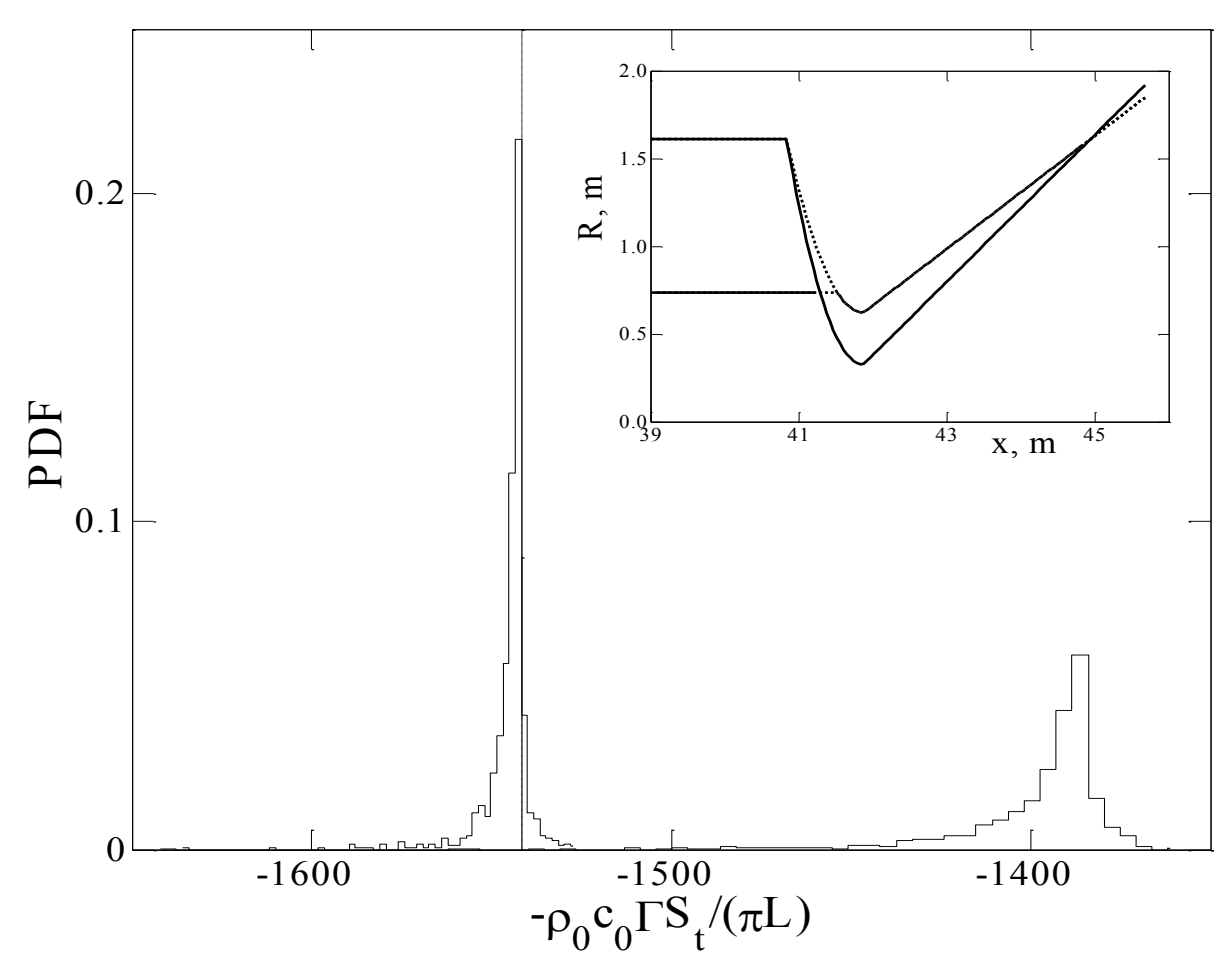

**Figure 3 Estimation of the value of the parameter -$c_0 \Gamma S_t/(\pi L)$ before (left curve) and after (right curve) the step-wise change of the nozzle throat area. The dashed line shows the actual value of the parameter. The solid lines show the PDF of the parameter estimation with $T$=0.1 sec, $\Delta t$=h=0.001 sec, $N$=500. The inset shows the sudden change in the nozzle throat geometry.**

Here the vector field is parameterized in special form [1] $f(x;c) = \hat{U}(x)c$, where $U(x)$ is a block-matrix with $N$ blocks of the form $\hat{I}\phi_n(x(t_k))$, $\hat{I}$ is $L$x$L$ unit matrix, and $\mathrm{v}_m(\mathbf{x}) = \sum_{n=1}^{N}\frac{\partial U_{nm}(\mathbf{x})}{\partial x_n}$.

| Parameter | Actual | Inferred | Relative error |
|---|---|---|---|
| $2\gamma\rho_p$ | 2575.34 | 2572 | 0.1% |
| $-c_0\Gamma S_t/(\pi L)$ | -1101.75 | -1103 | 1.1% |
| $D_{11}$ | 0.0902 | 0.0906 | 0.4% |
| $D_{22}$ | 0.0902 | 0.0906 | 0.4% |

Table 1 Results of the parameter estimation of the model (9) in the absence of the nozzle ablation. The total time of the measurements in this test was $T$=0.5 sec, the sampling time was $\Delta t$=0.001 sec, and the number of measured points was $N$=500.

The application of the algorithm (13)-(16) to the problem of inference of the abrupt changes of the gas flow parameters in the SRB due to the changes in the rocket geometry are presented in Figure 3 and Table 1. It can be seen from the figure and table that the algorithm can (at least in principle) infer sudden changes in the gas flow with time resolution of the order of 0.1 sec with relative error less than 1%.

Below it is assumed that the algorithm was applied to learn the SRB parameters in the nominal regime and multiple ground and flying tests. Therefore, nominal SRB parameters are considered to be known and we are mainly concerned with diagnostics and prognostics of the case breach fault parameters.

## 5. Example I: Self-consistent iterative algorithm

In this example we analyze the following problem. A hole

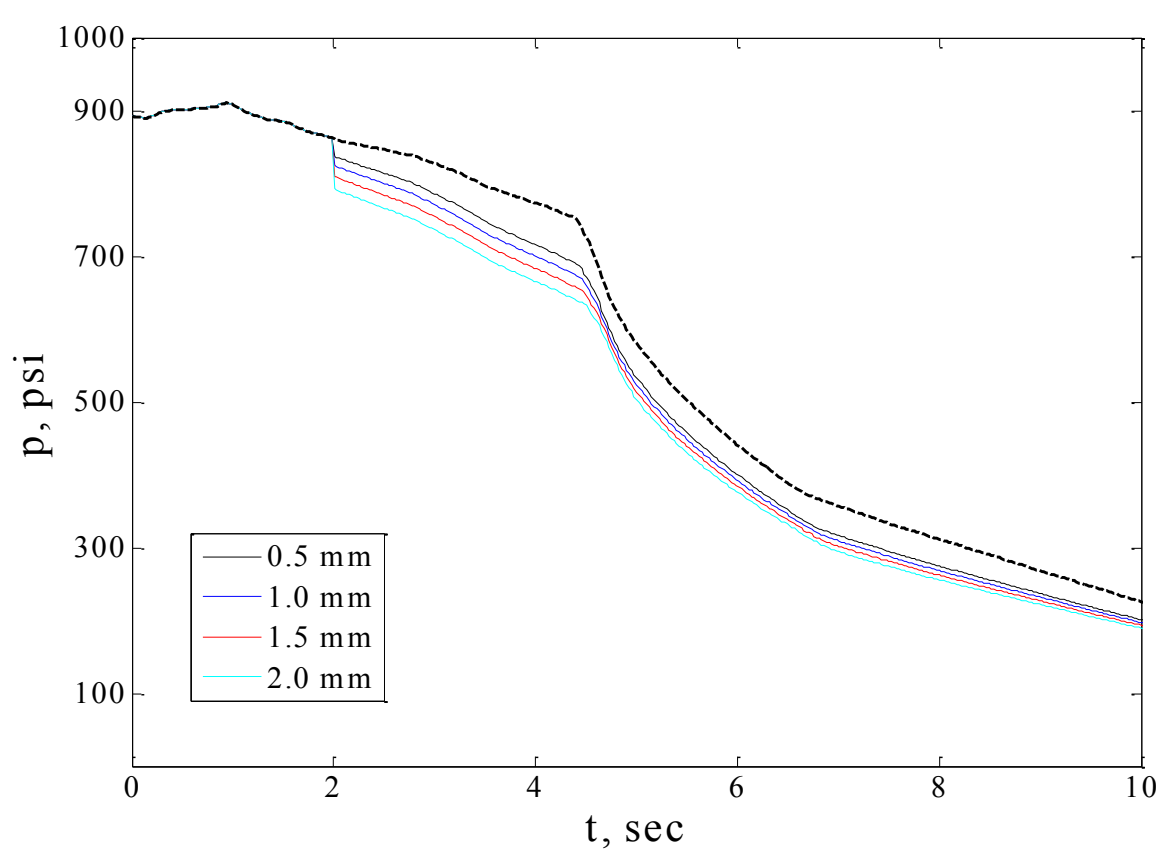

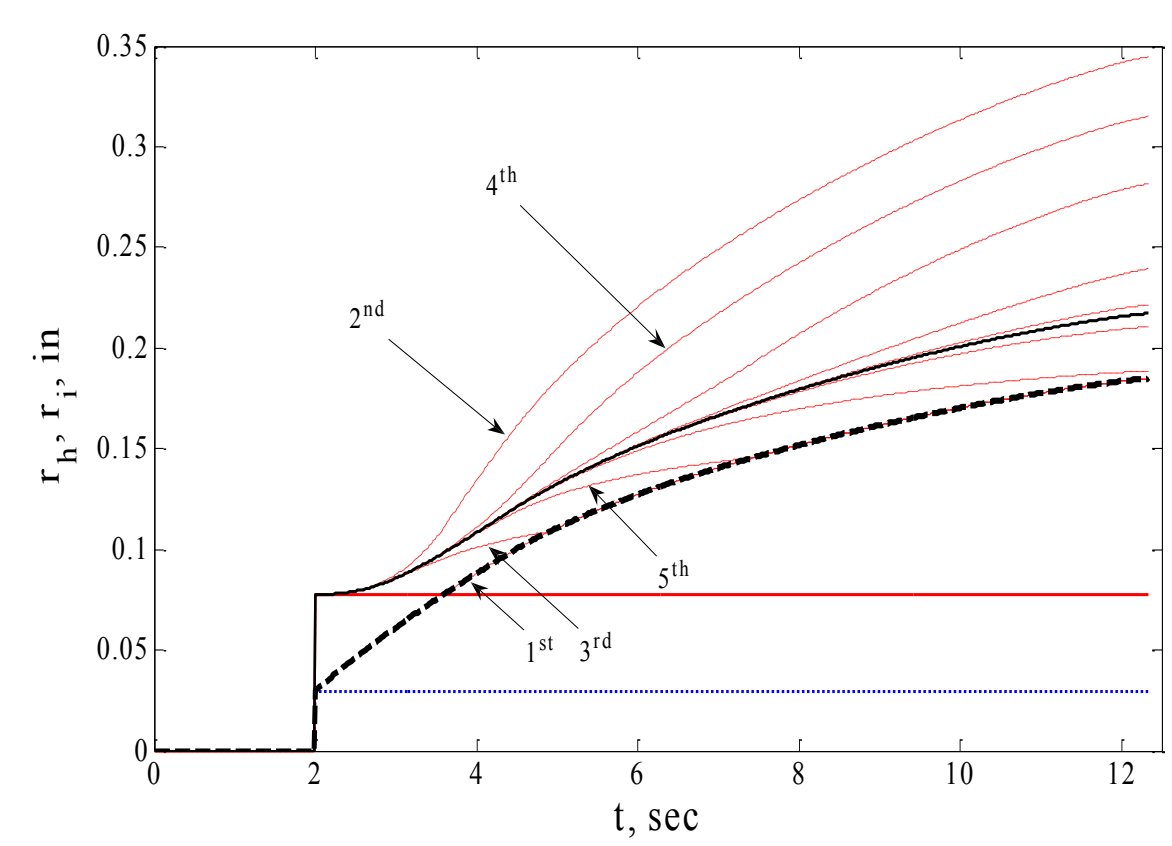

**Figure 4 (left) Results of the self-consistent calculations of the rocket parameters in the off-nominal regime using iterative algorithm A. Absolute values of pressure for four different initial values of the hole in the insulator: 0.5, 1.0, 1.5 and 2.0 mm are shown by the black, blue, red, and cyan solid lines respectively. The nominal pressure is shown by the dashed black line. (right) Iterations of the effective hole radius in the metal case. 0th approximation is shown by the red solid line. Five first approximations shown by red dashed lines are indicated by arrows. Final radius of the hole in the metal case is shown by black line. 0th approximation for the hole in the insulator is shown by dashed blue line. Final radius of the hole in the insulator is shown by the black dashed line.**

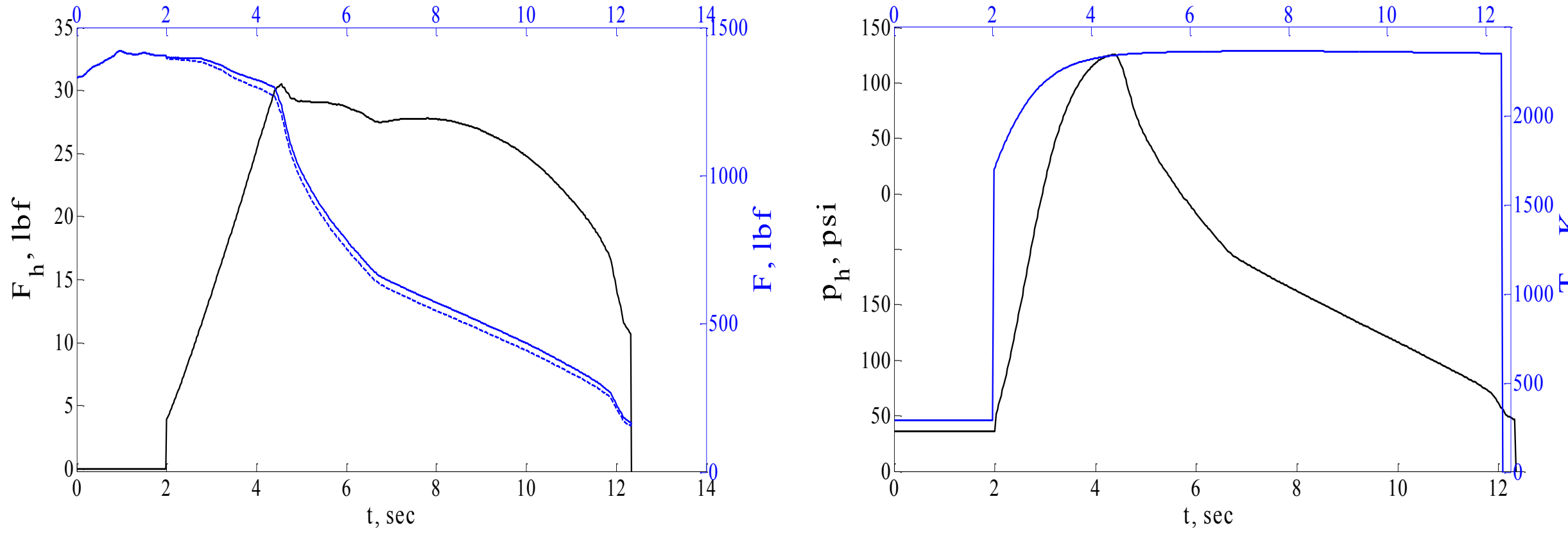

**Figure 5 (left) Fault-induced thrust (black solid line) is shown in comparison with nominal SRB thrust (blue solid line) and off-nominal SRB thrust (dashed blue line). Initial radius of the hole in the insulator is 0.75mm. (right) Pressure (black line) and temperature (blue line) in the metal hole through the case determined by the iteration algorithm B.**

through the metal case and insulator occurs suddenly at the initial time of the fault $t_0$. We have to infer and predict the dynamics of the growth of the holes in the insulator layer and in the metal case, as well as the fault-induced side thrust, and changes of the SRB thrust in the off-nominal regime. As an input, we use time-traces of the stagnation pressure in the nominal regime and nominal values of the SRB parameters. In particular, it is assumed that the ablation parameters for the nozzle and insulator materials and the melting parameters for the metal case are known. In this case we assume that the hole radius in the metal case is larger than the hole radius in the insulator (i.e. the velocity of the ablation of the insulator material is smaller than the velocity of the melting front), accordingly the fault dynamics is determined by the insulator ablation of the insulator (see details in [3], [14]).

To solve this problem we introduce a prognostics algorithm of the fault dynamics based on a self-consistent iterative algorithm that avoids numerical solution of the LDPM. We notice that with the limit of steady burning, the equations in (9), (11) can be integrated analytically. Because the hole throat is determined by the radius of the hole in the insulator layer, at this stage we can omit the equation for the hole radius in metal case. The resulting set of equations has the form

$$\begin{cases} p(t) = p_N(t)\left[\dfrac{S_{bef}(t)s_{tN}(t)}{S_{bN}s_{et}(t)}\right]^{\frac{1}{1-n}}, \\ R = a\int_o^t p^n(\tau)d\tau, \\ r_t^{1-\beta_t}(t) = r_{t0}^{1-\beta_t} + (1-\beta_t)v_{n0}\int_{t_0}^t p^{1-\beta_t}(\tau)d\tau, \\ r_i^{1-\beta_i}(t) = r_{i0}^{1-\beta_i} + \left(1-\beta_i\right)v_{i0}\int_{t_0}^t p^{1-\beta_i}(\tau)d\tau. \end{cases} \tag{17}$$

Here $s_{et}(t) = s_{tN}(t) + \Delta s_t(t) + s_h(t)$ is an effective nozzle throat area where the 1st term corresponds to the nominal regime, the 2nd corresponds to the deviation of the nozzle throat area from the nominal regime due to the fault, and the 3rd term corresponds to the area of the hole throat in the rocket case. Similarly, we define the effective burning area $S_{bet}(t) = S_{bN}(t) + \Delta S_b(t)$ as a sum of the burning area in the nominal regime and a term that describes the deviation of the burning area from the nominal regime due to the fault. Using Eqs (17) the following iterative algorithm **A1** can be introduced:

(1) Set initial values of the corrections to the nozzle and burning area to zero $\Delta s_t(t) = 0$ and $\Delta S_b(t)$=0. Set values of areas of the holes in the insulator and in the metal to constant initial values $s_h(t) = \pi \cdot r_{h0}^2$ and $s_{met}(t) = \pi \cdot r_{met0}^2$.
(2) Update time-trace of the pressure using 1st Eq. in (17)
(3) Update burn web distance $R$, radius of the hole in the insulator $r_h$, and nozzle throat radius $r_t$.
(4) Repeat from the step (2) until convergence is reached.

The results of the application of this self-consistent algorithm to the prognostics of the case breach fault parameters are shown in Figure 4 (left). Once quasi-steady pressure and the dynamics of growth of the hole in the insulator are predicted in the off-nominal regime one can determine the dynamics of the hole growth in the metal case and the dynamics of the fault-induced side thrust. To do so, we use the following self-consistent iterative algorithm **B** for $t>t_0$ that takes into account the assumption that the velocity of the melting front is larger than the velocity of ablation in the insulator

(1) Set 0th approximation $r^{(0)}_{met0}(t)$ for the hole radius in the metal to $r_{met0}$.
(2) Construct 1st approximation

$$r_h^{(1)}(t) = \begin{cases} r_h^{(0)}(t), & \text{if } r_h^{(0)}(t) < r_{met}(t) \\ r_{met}(t), & \text{if } r_h^{(0)}(t) \geq r_{met}(t) \end{cases}$$

(3) Solve Mach equation (use a maximum root)

$$M_0(t) = \begin{cases} M_0\left(1-\frac{\gamma-1}{2}M_0^2\right)^{\frac{1}{\gamma-1}} = \frac{s_i}{\Gamma s_h}, & \text{if } r_h^{(0)}(t) < r_{met}(t) \\ \frac{c_t}{c_0} = \sqrt{\frac{2}{\gamma+1}}, & \text{if } r_h^{(0)}(t) \geq r_{met}(t) \end{cases}$$

(4) Find the effective radius of the hole in the metal case

$$r_{met}^{1+\beta} = r_{met0}^{1+\beta} + \nu_{f0}\int_{t_0}^{t}\left(\frac{(\rho v)_h(\tau)}{(\rho v)_*^m}\right)^{1-\beta}\left(\frac{T_{met,h}(\tau)-T_{m0}}{T_*-T_{m0}}\right)d\tau$$

(5) Use Eqs (8) to find velocity $u_h(t)=M_{0h}(t)\cdot c_0$, mass flow $(\rho u)_h(t)$, temperature $T_h(t)$, and pressure $p_h(t)$ in the metal hole;
(6) Repeat from the step (2) until convergence is reached.
(7) Calculate fault-induced side thrust

$$F_h = \left[(\rho u)_h u_{h,ex} + (p_{h,ex} - p_a)\right]s_{met}.$$

A similar algorithm is used to find the ablation of the nozzle and SRB thrust in the off-nominal regime in the case breach fault. The results of these calculations are shown in Figure 4 (right) and Figure 5. We note that the inference of the fault parameters is trivially achieved using the same iterative algorithm with the only exception being that the time-trace for pressure in the off-nominal regime is given by the measurements. Accordingly the first equation in the set (17) and the 2nd step in the iterative algorithm **A** are not needed. Note also that the important part of the introduced above algorithms is an assumption that the design curve $S_b= f(R)$ representing the relation between the burning area $S_b$ and burn web distance $R$ is know and remains invariant characteristics of the SRB in the off-nominal regime of the case breach fault.

The deep physical meaning of the iterative procedure introduced above is that the ablation of the nozzle and the case breach fault develop in a self-consistent manner. Indeed, the increase of the leak cross section, due to insulator ablation under the action of the hot gas flow, leads to decreased pressure and hence a decreased burning rate. This, in turn, decreases the hot gas flow through the hole and the ablation rate. In this way a quasi-stationary regime of burning and ablation is developed. The parameters of burning in this regime can be found in a self-consistent way using an iterative algorithm, without integration of the full system of differential equations of motion.

## 6. Example II: Case Breach Fault with Nontrivial Geometry

In this Section we consider a situation, motivated by recent ground tests, in which fault occurs in a closure holding a pressure sensor. The disk has a form of a metal disk of radius $R_d$ and is covered by insulator material. We assume that a small leak arises on a boundary of the disk at moment $t=t_0$ (see Figure 6). Subsequently the small hole will grow due to surface melting and flame erosion (burning) of the metal under the action of the hot gas flow effluent from the combustion chamber. The problem of the FD&P is to detect fault, infer a model of the fault dynamics, and to predict pressure and thrust dynamics ahead of time in the presence of the fault. At the input of the algorithm we have time-traces of pressure and thrust, and known nominal parameters of the SRB and ablation of the nozzle. The difficulty of this problem is related to the fact that in the presence of the complicated geometry of the fault it is not sufficient to know the melting and burning parameters of the metal case. Indeed, in this case the last equation in (9) has to be substituted with the following set of equations

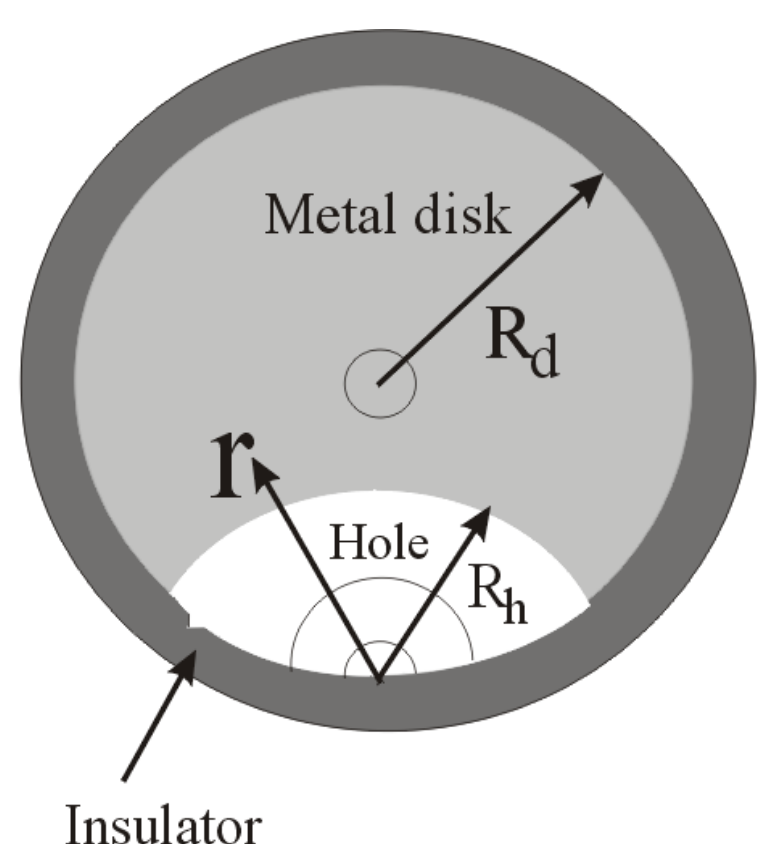

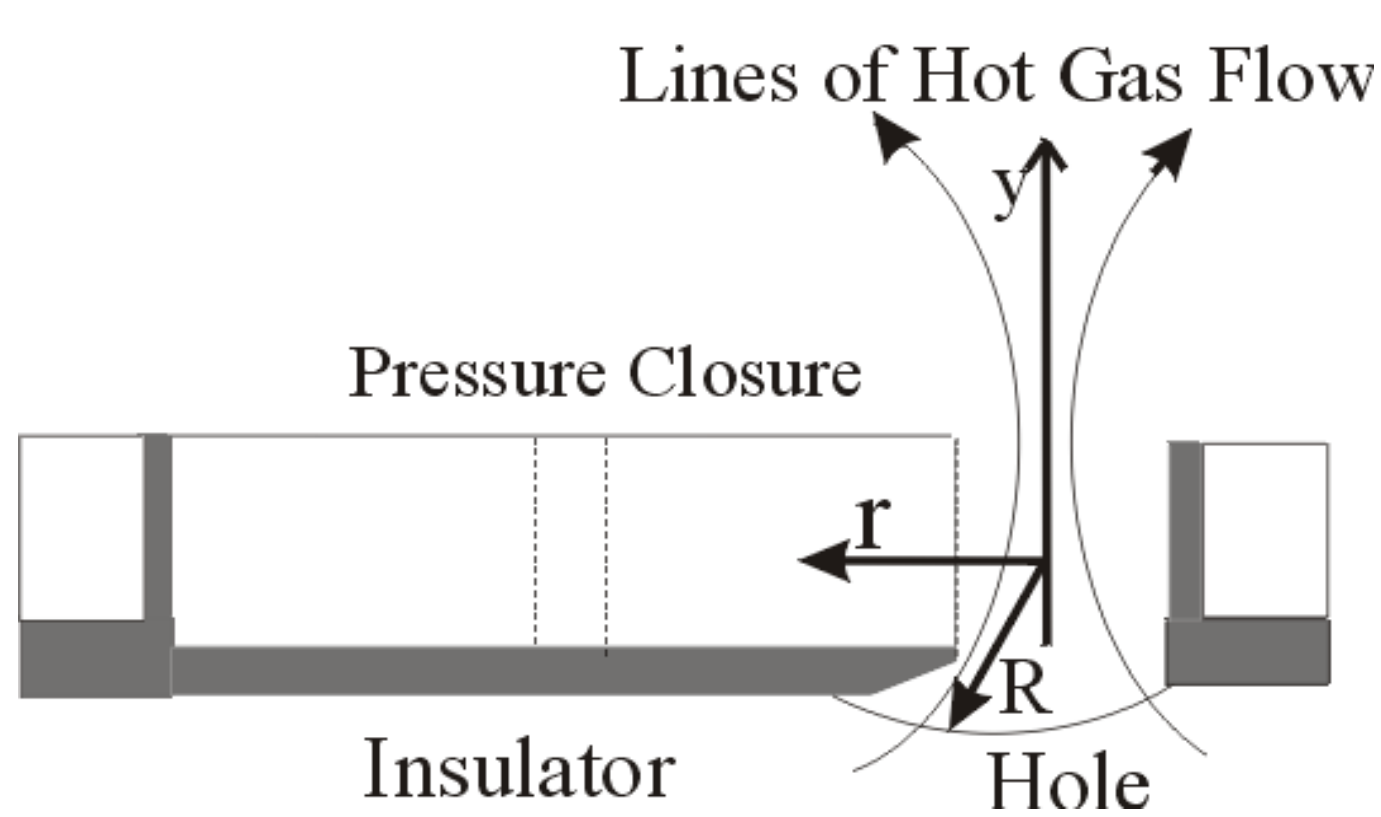

**Figure 6 Sketch of an idealized model (left) of hole growth and distribution (right) of hot gas flow near the hole.**

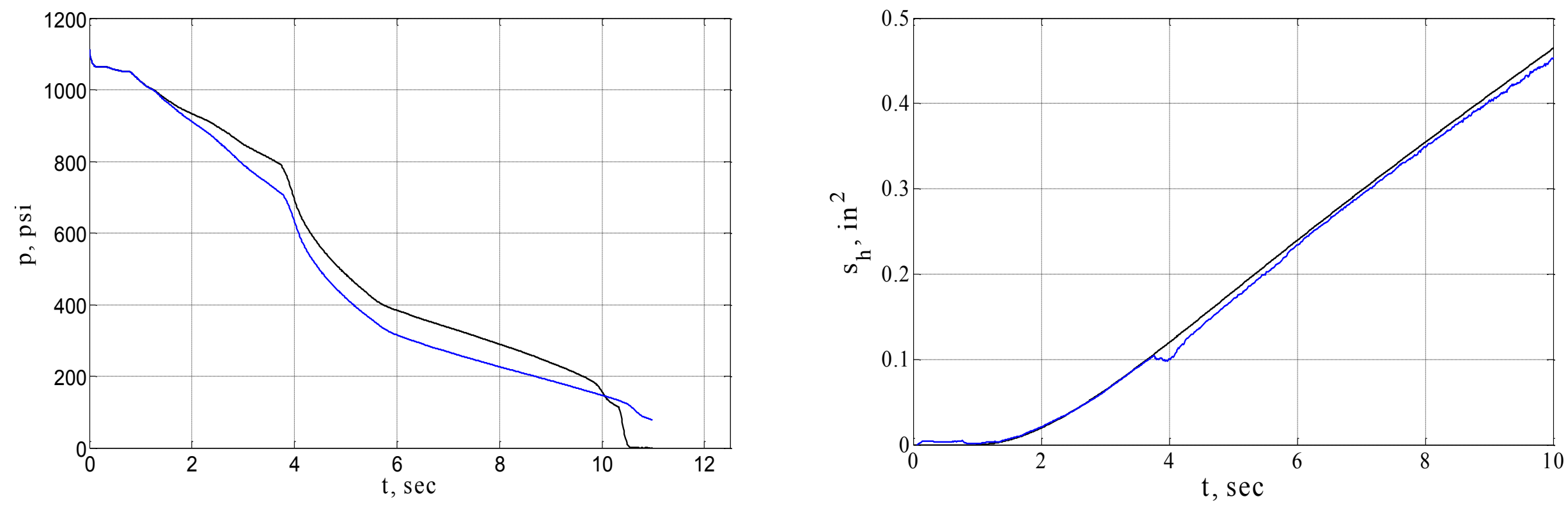

**Figure 7 (left) Pressure time-trace $p(t)$ in case breach fault-induced off-nominal regime (blue line) as compared to $p(t)$ in the nominal regime (black line). (right) The result of the reconstruction of the time-trace of the hole area $s_h(t)$ (blue line) using algorithm C as compared to the actual dynamics of the fault (black line).**

$$\dot{r}_h = \frac{Q_c + Q_R + Q_b}{\left[q_m + C_m\left(T_{mel} - T_{m0}\right)\right]\rho_m} = \frac{Q_c + Q_R}{\left[q_m + C_m\left(T_{mel} - T_{m0}\right)\right]\rho_m} + v_{fb}, \qquad (18)$$

$$s_h = R_d^2\left[\pi - \left(2 - x^2\right)\arccos\left(\frac{x}{2}\right) - x\sqrt{1 - \frac{x^2}{4}}\right].$$

Here we take into account the geometry of the fault and the heating of the disk due to the convective and radiation flows with the values of $Q_R$ and $Q_c$ given by equations (29) and (32) and $v_{fb}$ is a constant rate of erosive burning (see Appendix). Note that to find ablation of the nozzle exit we can use time-traces of $p(t)$ and $r_t(t)$ obtained by integration of 3 equation in set of Eqs. (9) and the equation for the ablation of the nozzle exit

$$\dot{r}_n = v_{n0}\left(\frac{p r_t^2}{r_n^2}\right)^{1-\beta_t}\left(\frac{r_n}{r_{n0}}\right)^{-\beta_t}\left(\frac{T_n - T_a}{T_* - T_a}\right) \qquad (19)$$

where $T_n$ is given by the solution of the last two equations in (8) for a given $S_t = \pi r_n^2$.

Note that the relation between hole radius $r_h$ and hole area $s_h$ was also changed to take into account the complex geometry of the fault. Accordingly, the fault dynamics becomes highly nonlinear. This is true even in the case of metal melting with constant so-called erosion velocity $v_{er}$ of the order of 0.1 in/sec, which is frequently observed in experiments (see e.g. [29]). The reason is that there exists a delay time (related to the heating of the metal) before a steady regime of melting can be established. As a consequence the typical time-trace of the pressure in the off-nominal regime may have the form shown in the Figure 7 (left) by the blue line. In this case the hole in the case appears at $t_0$ = 1 sec and grows with a velocity of approximately 0.16 in/sec.

To infer parameters of the fault we introduce the following algorithm **C**

(1) Given the pressure time-traces $p(t)$, integrate equations for the nozzle ablation (equation 5 in (9) and equation (19)) to obtain $s_t(t)$ and $s_{ex}(t)$ in the off-nominal regime.

(2) Use equation 3 in (9) and $p(t)$ to obtain time-trace of the web distance $R(t)$

(3) Use $R(t)$ and the design curve $S_b = f(R)$ to find time-trace of the burning area $S_b(t)$

(4) Use $S_b(t)$ and quasi-stationary solution of the equation for the pressure to find a time trace of the effective area of the nozzle throat

$$s_{et}(t) = \left(\frac{S_b(t)}{p(t)^{1-n}}\right)p_{00}, \quad p_{00} = \frac{a c_0 \rho_p \rho_0 \Gamma}{\gamma}.$$

(5) Find time-trace of the hole throat area $s_h(t) = s_{et}(t) - s_t(t)$

(6) Find nozzle thrust and fault-induced thrust.

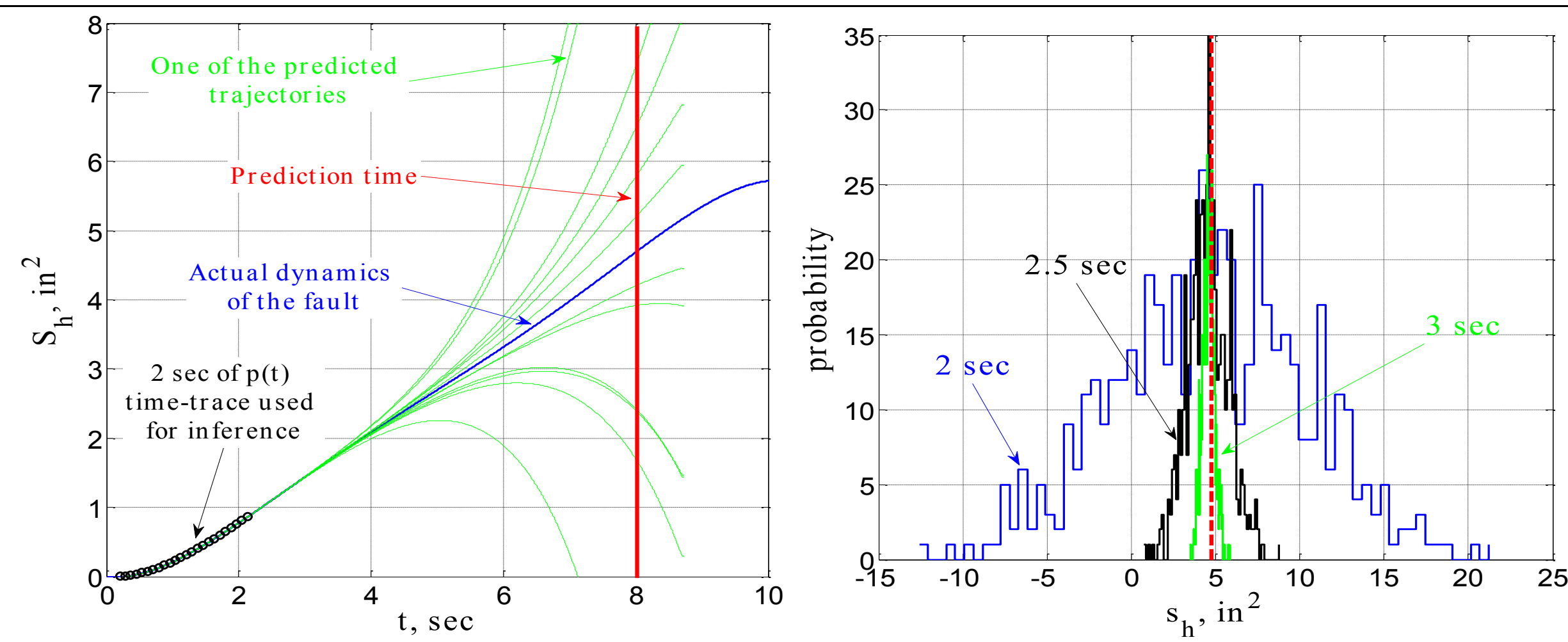

**Figure 9 (left) Actual fault dynamics is shown by the blue solid line. The time interval elapsed from the case breach fault is shown by the black circles. The predicted trajectories of the fault are shown by thing green solid lines. The time moment of the prediction is indicated by a vertical red line. (right) The distribution of the predicted value of the fault at the future time t = 8 sec is shown by solid blue, black, and green lines in comparison with the true value of the fault indicated by the vertical red dashed line. The time elapsed from the case breach fault used for predictions is shown in the figure.**

The result of the application of this algorithm to the analysis of the time-trace of the pressure is shown in Figure 7 (right) and Figure 8. It can be seen from the figures that the algorithm C can infer accurately the fault dynamics. We now apply these results to develop an algorithm of prognostics.

## 7. PROGNOSTICS ALGORITHM FOR COMPLEX GEOMETRY

First we note that prognostics algorithms are much more sensitive towards the errors of parameter estimations than the diagnostics algorithms. For example a simple polynomial fit of the first 2.5 sec of the inferred time-trace of $s_h(t)$ shown in Figure 7 (left) is not sufficient to predict fault dynamics. The corresponding predictions will quickly diverge from the actual value of $s_h(t) \approx 2$ sec later. Therefore, more sophisticated algorithms are required to predict nonlinear fault dynamics. Such algorithms can be developed using the LDPM (9) and Bayesian dynamical inference discussed briefly in Sec. 4.

To apply the dynamical inference algorithm to the prognostics of the case breach fault we notice that metal erosion of the hole walls during the case breach fault is mainly determined by the constant rate $v_b$ (see e.g. [29], model (18), and the results of simulations shown in Figure 7 (right)). And if the fault geometry is simple, which corresponds to the hole in the side wall, the $s_h(t) \propto t$, and the predictions will be considerably simplified. If, however, the hole occurs at one of the joins, which is very likely the case in practice, the time evolution of the $s_h(t)$ will be mostly determined by the geometry of the joins and the predictions will be mainly based on the earlier detection of the catastrophic trends in the $s_h(t)$.

The catastrophic trends in the $s_h(t)$ are usually related to the high powers $t^m$ of $t$, because these terms determine fast deviations of the dynamics from the nominal regime. In this context it is important to verify convergence of the method that includes higher order terms into the set of the base

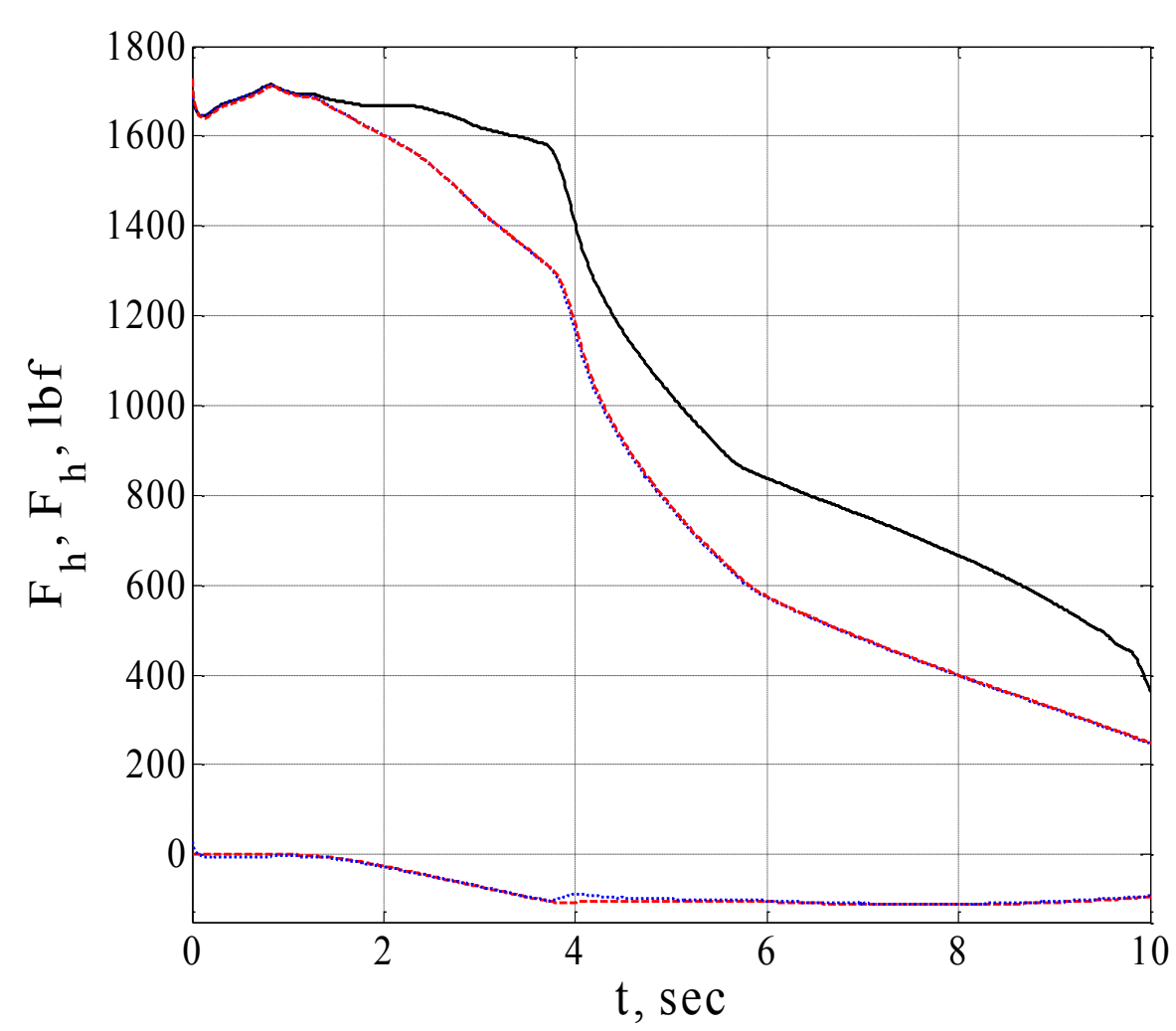

**Figure 8 Nominal pressure (black line) as compared with the off-nominal pressure of the nozzle (blue dashed line) and of the hole (blue dashed lines at the bottom). Inferred values of the nozzle and hole thrusts are shown by red dashed lines.**

functions of $s_h(t)$.

To model such situation we will consider practically important case when only pressure signals $p(t)$ (and perhaps temperature $T(t)$) are given by the measurements. Then the model (9) can be reduced to the following form

$$\dot{p} = -\frac{c_0 \gamma \Gamma s_{tf}}{V r_b} p\sqrt{T} + \frac{F}{V}\left(\gamma \rho_p - p\right) p^n + \sqrt{D}\xi_2(t), \quad (20)$$

Here $T = p/\rho \approx 1$ for dimensionless pressure and density introduced in (10), and the time traces $R(t)$, $F(t)$, $V(t)$, and $r_t(t)$ are completely determined by the pressure and inferred law of the nozzle ablation

$$R(t) = \int_0^t p^n(t')dt', \quad F(t) = f\left(R(t)\right),$$

$$V(t) = V_0 + \int_0^{R(t)} f(R)dR, \quad s_t(t) = \pi r_t^2(t), \qquad (21)$$

$$r_t\ (t) = \left[ r_{t0}^{1+\beta} + v_m(1+\beta)\int_0^t p^{1-\beta}(t')dt' \right]^{\frac{1}{1+\beta}},$$

and, therefore, are known.

We choose $s_h(t) = s_{et}(t) - s_t(t)$ in the form

$$s_h(t) = \sum_{m=1}^{5} a_m t^m, \qquad (22)$$

and apply dynamical inference algorithm introduce in Sec. 4 to infer parameters $a_i$ and $D$ in (20). All other parameters in (20) are know from the analysis of the nominal regime.

The 1D problem (20), (22) can be solved by applying Eqs (13)-(16), where $U(x_n)$ has the form

$$U = \left\{1, t, t^2, t^3, t^4, t^5\right\}. \qquad (23)$$

The results of the predictions are shown in the Figure 9. In the left figure we show the time elapsed from the fault ($t$ = 2 sec) by black circles. The pressure time-traces measured during the first 2 sec are then used to infer parameters $a_i$ and $D$. Next, parameters $a_i$ are used to predict the future time evolution of the fault using (22). The distribution of the predicted trajectories obtained in this way is shown by the blue line in the Figure 9 (right). When new data are measured on the interval 2.5 sec and 3 sec the procedure is repeated to obtain the updated distribution of the predicted values of the fault as shown in the Figure 9 (right) by black and green lines. It can be seen from the figure that the method converges during first 3 sec of the fault.

## 8. Summary and Discussion

In this paper we report the work in progress on the development of hybrid probabilistic algorithms for SRBs on-board IVHM system. Two experimentally motivated situations are considered. In the first situation case breach fault occurs suddenly in the metal case with simple geometry. In the second situation the fault develops slowly in the forward closure with complex geometry. A high-fidelity model of the fault dynamics is developed in FLUENT. The LDPM of the off-nominal regime induced by the case breach is derived using the high-fidelity model. A number of algorithms are derived to infer fault parameters and to predict fault dynamics. The algorithms accommodate abrupt changes in the model parameters and can be used to develop hybrid probabilistic on-board SRB IVHM. The performance of the algorithms was tested using analysis of the experimental time-series data. It is shown that the algorithm can be successfully applied for the prognostics of the case breach fault.

The developed methods and algorithms can be used to analyze other SBR' faults, including overpressure and breakage of the case induced by nozzle blocking, bore choking and grain deformation. The bore choking phenomenon is an almost radial deformation (bulge) of the propellant near booster joint segments. This phenomenon can cause choking of the exhaust gas flow and increase the burning surface which can lead to critical overpressure in the combustion chamber. Development testing has shown that this fault was observed, for example, in primary construction of the Titan IV (see the report [6], [7], [31]). The bore choking, and also cracks and voids in the solid propellant, can result in local burning of the booster case and also in abrupt breaking off of large pieces of the propellant. These pieces can stick in the nozzle throat and block the exhaust gas flow (nozzle blocking fault). In all these cases the fault dynamics is governed by the changes in the burning area and/or effective nozzle area. Therefore the LDPM introduced in this work can be efficiently applied to analyze of these faults and develop on-board SRB IVHM.

## Appendix :

The goal of the appendix is to study ablation of the nozzle surface and also melting and burning of the metal surface of a leak hole. These surfaces heat under the action of a hot gas stream $Q$ flowing through the hole or nozzle. The surfaces start to corrode when the surface temperature $T_s$ reaches the ablation temperature of the insulator material $T_{abl}$ or the melting point of the metal $T_{mel}$, respectively. Let us analyze the growth of $T_s$ on the surface of a hole of a radius $R_h$ in a metal disk or nozzle throat of a radius $R_N$. Temperature distribution inside the material is described by

$$C\rho \frac{\partial T}{\partial t} = \kappa \left( \frac{\partial^2 T}{\partial r^2} + \frac{1}{r}\frac{\partial T}{\partial r} \right) \qquad (24)$$

where $r$ is a coordinate normal to the surface at $r > R_{h,N}$. It is well known that a homogeneous solution of this equation is

$$T(x,t) = A(t)\exp\left[ -\frac{r^2}{4\gamma^2 t} \right] \quad where \quad \gamma^2 = \frac{\kappa}{c\rho} \qquad (25)$$

i. e. thermo-diffusion length is equal to:

$$l_D = 2\sqrt{\kappa t / c\rho} \qquad (26)$$

Here the parameters $C$, $\rho$, $\kappa$ correspond to a metal or an insulator material of a nozzle. The flow of the hot gas Q for the time interval $dt$ will heat the surface layer of an area $dS$ and a thickness of $l_D$ . For example, the heat balance for the metal surface can be written as:

$$QdSdt = c\rho dSd\left[(T_S - T_{m0})l_D\right] \tag{27}$$

where $T_{m0}$ is an initial temperature of the metal disk. It follows from Eqs. (26) and (27)

$$\frac{dT_S}{dt} + \frac{\left(T_S - T_{m0}\right)}{2t} = \frac{Q}{2\sqrt{C\rho\kappa t}} \tag{28}$$

Here we assume $Q = Q_h = Q_c + Q_R$ where $Q_R$ and $Q_C$ are radiation and convective heat flows from the hot gas to the metal surface. Radiation heat flow is given by

$$Q_R = \sigma\left[1 - \exp\left(-\lambda P\right)\right]\left(T^4 - T_{m0}^4\right) \tag{29}$$

where $\lambda = \left[0.001 + 4\times10^{-4} \times \%AL\right]/14.69/psi$ is the emissivity of the hot gas, $\%AL$ is the percentage of aluminum in a solid propellant, and $\sigma = 5.67\times10^{-8} W/m^2K^4$ is the Stefan-Boltzmann constant. For realistic parameters $\lambda P << 1$, therefore $Q_R \approx \sigma\lambda P\left(T^4 - T_{m0}^4\right)$. The convective heat flow $Q_C = h_g(T - T_S)$. When the surface temperature $T_s$ reaches to $T_{mel}$, the value of $T_s$ is equal to $T_{mel}$ inasmuch as the melted metal layer rushes out from the hole almost immediately for a time t<1msec [14]. The heat transfer coefficient $h_g$ can be written as [20]-[23].

$$h = 0.5JC_p\eta_r C_f \tag{30}$$

Where the coefficient $\eta_r$ describes an effect of the roughness. The value of the coefficient $C_f$ is determined by Reynolds number: $\mathrm{Re} = \rho ux/\mu$. $Re \approx 10^6$ for our parameters and p=1000psi. The critical Reynolds number which determines a transition between laminar and turbulent regimes is known to vary from approximately $10^5$ to $3\times10^6$, depending on the surface roughness and the turbulence level of the free stream [20]-[21]. The turbulence regime probably occurs even for $p \approx 100psi$ due to growth of the hole radius and the roughness of the metal surface. $C_f$ can be approximated over a fairly wide range of Reynolds numbers by [20]-[23]

$$C_f = 0.046\left(\frac{jD^*}{\mu}\right)^{-0.2}\left(\frac{\mu C_p}{\kappa}\right) = 0.046\,\mathrm{Re}^{-0.2}\,\mathrm{Pr}^{-0.67} \tag{31}$$

Here the Prandtl number $\mathrm{Pr} = \mu C_p/\kappa \approx 1$ [19]-[21]. It follows from Eqs. (30) and (31)

$$Q_C = 0.023\eta_r C_p\left(\frac{\gamma p}{\Gamma c_0}\right)^{0.8}\left(\frac{2R}{\mu}\right)^{-0.2}(T^* - T_S) \tag{32}$$

where $T^* = 2T/\left(1+\gamma\right)$. The increase of the surface temperature is given by Eq. (28) where $Q_c$ and $Q_R$ are given by Eqs. (32) and (29). Eq. (28) determines a time $t=t_0$ when the surface temperature $T_s$ reaches $T_m$. Our calculations show that $t_0<0.2sec$ for a steel and $t_0<<0.1sec$ for a typical nozzle material.

The melting starts at time t>$t_0$ when $T_s=T_{mel}$. The velocity of melting can be found from the heat balance equation:

$$QdSdt = [q + c(T - T_0)]\rho dSv_f dl_D \tag{33}$$

where $T_0$ is an initial temperature. It follows that the velocity of melting front for time t>$t_0$ is given by

$$\dot{h}_{met} = v_f = \frac{Q_c + Q_R}{\left[q_m + C_m\left(T_{mel} - T_0\right)\right]\rho_m} \tag{34}$$

The metal surface can react with oxidizing agents of the combustion products (hot gas). Obviously, velocity of propagation of metal burning front $v_{fb}$ has to become saturated at high pressures and/or intense surface flows of oxidizing agents. Experiments show that $v_{fb} \approx 0.76cm/\sec$ for iron and carbon steels [21], [22]. The same value of $v_{fb}$ was reported in the study of ATK Thiokon Inc. team [29]. We notice that the gas temperature T=T* inside the hole is greater than temperature of the metal surface, therefore a heat flow $Q_b$ induced by the burning spends for the metal melting. Taking into account the burning of the metal surface we can written $v_f$ in the following form

$$\dot{h}_h = v_f = \frac{Q_c + Q_R + Q_b}{\left[q_m + C_m\left(T_{mel} - T_{m0}\right)\right]\rho_m} = \frac{Q_c + Q_R}{\left[q_m + C_m\left(T_{mel} - T_{m0}\right)\right]\rho_m} + v_{fb} \tag{35}$$

Analogously Eq. (34), the velocity of insulator material ablation of nozzle t>$t_0$ is given by

$$\dot{h} = v_{abl} = \frac{Q_c + Q_R}{\left[q_{in} + C_{in}\left(T_{abl} - T_0\right)\right]\rho_{in}} \tag{36}$$

Estimations show that $Q_c$ is much greater than both $Q_R$ Thus, we can use Eqs. (4) and (5) where $v_n$ and $v_m$ are approximately equal to

$$v_n = \frac{2.3(T^* - T_{abl})}{100} \eta_r C_p \left(\frac{\gamma p_0}{\Gamma c_0}\right)^{0.8} \left(\frac{2r_{n0}}{\mu}\right)^{-0.2},$$

$$v_m = \frac{2.3(T^* - T_{mel})}{100} \eta_r C_p \left(\frac{\gamma p_0}{\Gamma c_0}\right)^{0.8} \left(\frac{2r_{h0}}{\mu}\right)^{-0.2}$$